\documentclass[12pt]{article}
\usepackage{psfig,mydef}
\begin{document}
\begin{titlepage}
\mbox{} \vskip 3cm \centerline{\LARGE\bf \sf Information Theory in Molecular
Biology} \vskip 1in \centerline{\large \bf \sf Christoph Adami$^{1,2}$} \vskip
0.2in \centerline{\it $^1$Keck Graduate Institute of Applied Life Sciences,}
\centerline{\it 535 Watson Drive, Claremont, CA 91711}\vskip 0.5cm \centerline{\it
$^2$Digital Life Laboratory 139-74,}  \centerline {\it California Institute of
Technology, Pasadena, CA 91125} \vskip 2cm

\date{\today}
\centerline{\bf Abstract} This article introduces the physics of
information in the context of molecular biology and genomics.
Entropy and information, the two central concepts of Shannon's
theory of information and communication, are often confused with
each other but play transparent roles when applied to statistical
{\em ensembles} (i.e., identically prepared sets) of symbolic
sequences. Such an approach can distinguish between entropy and
information in genes, predict the secondary structure of
ribozymes, and detect the covariation between residues in folded
proteins. We also review applications to molecular sequence and
structure analysis, and introduce new tools in the
characterization of resistance mutations, and in drug design.

\vskip 4 cm
\end{titlepage}
\vskip 0.25cm In a curious twist of history, the dawn of the age
of genomics has both seen the rise of the science of
bioinformatics as a tool to cope with the enormous amounts of data
being generated daily, and the decline of the {\em theory} of
information as applied to molecular biology. Hailed as a harbinger
of a ``new movement''\nocite{quast53} (Quastler 1953) along with
Cybernetics, the principles of information theory were thought to
be applicable to the higher functions of living organisms, and
able to analyze such functions as metabolism, growth, and
differentiation (Quastler 1953). Today, the metaphors and the
jargon of information theory are still widely used (Maynard Smith
1999a, 1999b)\nocite{maynard99a,maynard99b}, as opposed to the
mathematical formalism, which is too often considered to be
inapplicable to biological information.

Clearly, looking back it appears that too much hope was laid upon
this theory's relevance for biology. However, there was
well-founded optimism that information theory ought to be able to
address the complex issues associated with the storage of
information in the genetic code, only to be repeatedly questioned
and rebuked (see, e.g., \nocite{vincent94,sarkar96}Vincent 1994,
Sarkar 1996). In this article, I outline the concepts of entropy
and information (as defined by Shannon) in the context of
molecular biology. We shall see that not only are these terms
well-defined and useful, they also coincide precisely with what we
intuitively mean when we speak about information stored in genes,
for example. I then present examples of applications of the theory
to measuring the information content of biomolecules, the
identification of polymorphisms, RNA and protein secondary
structure prediction, the prediction and analysis of molecular
interactions, and drug design.

\section{Entropy and Information}
Entropy and information are often used in conflicting manners in
the literature. A precise understanding, both mathematical and
intuitive, of the notion of information (and its relationship to
entropy) is crucial for applications in molecular biology.
Therefore, let us begin by outlining Shannon's original entropy
concept (Shannon, 1948).

\subsection{Shannon's Uncertainty Measure}
Entropy in Shannon's theory (defined mathematically below) is a
measure of uncertainty about the identity of objects in an
ensemble. Thus, while ``entropy'' and ``uncertainty'' can be used
interchangeably, they can {\it never} mean information. There is a
simple relationship between the entropy concept in information
theory and the Boltzmann-Gibbs entropy concept in thermodynamics,
briefly pointed out below.

Shannon entropy or uncertainty is usually defined with respect to
a particular observer.  More precisely, the entropy of a system
represents the amount of uncertainty {\em one particular observer}
has about the state of this system. The simplest example of a
system is a {\em random variable}, a mathematical object that can
be thought of as an $N$-sided die that is uneven, i.e., the
probability of it landing in any of its $N$ states is not equal
for all $N$ states. For our purposes, we can conveniently think of
a polymer of fixed length (fixed number of monomers), which can
take on any one of $N$ possible states, where each possible
sequence corresponds to one possible state. Thus, for a sequence
made of $L$ monomers taken from an alphabet of size $D$, we would
have $N=D^L$. The uncertainty we calculate below then describes
the observer's uncertainty about the true identity of the molecule
(among a very large number of identically prepared molecules: an
{\em ensemble}), given that he only has a certain amount of
probabilistic knowledge, as explained below.

This hypothetical molecule plays the role of a random variable if we
are given its {\em probability distribution}: the set of probabilities
$p_1,...,p_N$ to find it in its $N$ possible states.  Let us thus
call our random variable (random molecule) ``$X$'', and give the names
$x_1,...,x_N$ to its $N$ states. If $X$ will be found in state $x_i$
with probability $p_i$, then the entropy $H$ of $X$ is given by
Shannon's formula
\begin{equation}
H(X)=-\sum_{i=1}^N p_i\log p_i\;. \label{entropy}
\end{equation}
I have not here specified the basis of the log to be taken in the above formula.
Specifying it assigns units to the uncertainty. It is sometimes convenient to use
the number of possible states of $X$ as the base of the logarithm (in which case
the entropy is between zero and one), in other cases base 2 is convenient (leading
to an entropy in units ``bits''). For biomolecular sequences, a convenient unit
obtains by taking logarithms to the basis of the alphabet size, leading to an
entropy whose units we shall call ``mers". Then, the maximal entropy equals the
length of the sequence in mers.

Let us examine Eq.~(\ref{entropy}) more closely. If measured in
bits, a standard interpretation of $H(X)$ as an uncertainty
function connects it to the smallest number of ``yes-no" questions
necessary, on average, to identify the state of random variable
$X$. Because this series of yes/no questions can be thought of as
a {\em description} of the random variable, the entropy $H(X)$ can
also be viewed as the {\em length of the shortest description of
$X$} (Cover and Thomas, 1991). In case nothing is known about $X$,
this entropy is $H(X)=\log N$, the maximal value that $H(X)$ can
take on. This occurs if all states are equally likely:
$p_i=1/N\,;i=1,...,N$. If something (beyond the possible number of
states $N$) is known about $X$, this reduces our necessary number
of questions, or the length of tape necessary to describe $X$.
If I know that state $X=x_7$, for example, is highly
unlikely, then my uncertainty about $X$ is going to be smaller.

How do we ever learn anything about a system?  There are two choices. Either we
obtain the probability distribution using {\em prior knowledge} (for example, by
taking the system apart and predicting its states theoretically) or by making
measurements on it, which for example might reveal that not all states, in fact,
are taken on with the same probability. In both cases, the difference between the
maximal entropy and the remaining entropy after we have either done our
measurements or examined the system, is the amount of information we have about the
system. Before I write this into a formula, let me remark that, by its very
definition, information is a {\it relative} quantity. It measures the {\it
difference of uncertainty}, in the previous case the entropy before and after the
measurement, and thus can never be absolute, in the same sense as potential energy
in physics is not absolute. In fact, it is not a bad analogy to refer to entropy as
``potential information'', because potentially all of a system's entropy can be
transformed into information (for example by measurement).

\subsection{Information}
In the above case, information was the difference between the maximal and the
actual entropy of a system. This is not the most general definition as I have
alluded to. More generally, information measures the amount of {\it correlation}
between two systems, and reduces to a difference in entropies in special cases. To
define information properly, let me introduce another random variable or molecule
(call it ``$Y$''), which can be in states $y_1,...,y_M$ with probabilities
$p_1,...,p_M$. We can then, along with the entropy $H(Y)$, introduce the joint
entropy $H(XY)$, which measures my uncertainty about the joint system $XY$ (which
can be in $N\cdot M$ states). If $X$ and $Y$ are {\em independent} random variables
(like, e.g., two dice that are thrown independently) the joint entropy will be just
the sum of the entropy of each of the random variables. Not so if $X$ and $Y$ are
somehow connected.  Imagine, for example, two coins that are glued together at one
face. Then, heads for one of the coins will always imply tails for the other, and
vice versa. By gluing them together, the two coins can only take on two states, not
four, and the joint entropy is equal to the entropy of one of the coins.

The same is true for two molecules that can bind to each other.
First, remark that random molecules do not bind. Second, binding
is effected by mutual specificity, which requires that part of the
sequence of one of the molecules is interacting with the sequence
of the other, so that the joint entropy of the pair is much less
than the sum of entropies of each. Quite clearly, this binding
introduces strong correlations between the states of $X$ and $Y$:
if I know the state of one, I can make strong predictions about
the state of the other. The information that one molecule has {\em
about} the other is given by
\begin{equation}
I(X:Y)=H(X)+H(Y)-H(XY)\;, \label{info}
\end{equation}
i.e., it is the difference between the sum of the entropies of each, and the joint
entropy. The colon between $X$ and $Y$ in the notation for the information is
standard; it is supposed to remind the reader that information is a symmetric
quantity: what $X$ knows about $Y$, $Y$ also knows about $X$. For later reference,
let me introduce some more jargon. When more than one random variable is involved,
we can define the concept of {\em conditional entropy}. This is straightforward.
The entropy of $X$ conditional on $Y$ is the entropy of $X$ {\em given} $Y$, that
is, if I know which state $Y$ is in. It is denoted by $H(X|Y)$ (read ``$H$ of $X$
given $Y$'') and is calculated as
\begin{equation}
H(X|Y)=H(XY)-H(Y)\;.
\end{equation}
This formula is self-explanatory: the uncertainty I have about $X$ if $Y$ is known
is the uncertainty about the joint system minus the uncertainty about $Y$ alone.
The latter, namely the entropy of $Y$ without regard to $X$ (as opposed to
``conditional on $X$'') is sometimes called a {\em marginal} entropy. Using the
concept of conditional entropy, we can rewrite Eq.~(\ref{info}) as
\begin{equation}
I(X:Y)=H(X)-H(X|Y)\;.
\end{equation}

We have seen earlier that for independent variables
$H(XY)=H(X)+H(Y)$, so information measures the {\em deviation}
from independence. In fact, it measures exactly the amount by
which the entropy of $X$ or $Y$ is reduced by knowing the other,
$Y$ or $X$.  If $I$ is non-zero, knowing one of the molecules
allows you to make more accurate predictions about the other:
quite clearly this is exactly what we mean by information in
ordinary language. Note that this definition reduces to the
example given earlier (information as difference between
entropies), if the only possible correlations are {\em between}
$X$ and $Y$, while in the absence of the other each molecule is
equiprobable (meaning that any sequence is equally likely).  In
that case, the marginal entropy $H(X)$ must be maximal ($H=\log
N$) and the information is the difference between maximal and
actual (i.e., conditional) entropy, as before.

\subsection{Entropy in Thermodynamics}
I will briefly comment about the relationship between Shannon's theory and
thermodynamics (Adami and Cerf 1999)\nocite{adami99}. For the present purpose it
should suffice to remark that Boltzmann-Gibbs thermodynamic entropy is just like
Shannon entropy, only that the probability distribution $p_i$ is given by the
Boltzmann distribution of the relevant degrees of freedom (position and momentum):
\begin{equation}
\rho(p,q)= \frac1Ze^{-E(p,q)/kT}\;,
\end{equation}
and the thermodynamic quantity is made dimensional by multiplying
Shannon's dimensionless uncertainty by Boltzmann's constant. It
should not worry us that the degrees of freedom in thermodynamics
are continuous, because any particular measurement device that is
used to measure these quantities will have a finite resolution,
rendering these variables effectively discrete through
coarse-graining. More importantly, equilibrium thermodynamics
assumes that all entropies of isolated systems are at their
maximum, so there are no correlations in equilibrium thermodynamic
systems, and therefore there is {\em no information}. This is
important for our purposes, because it implies, a fortiori, that
the information stored in biological genomes guarantees that
living systems are far away from thermodynamical equilibrium.
Information theory can thus be viewed as a type of non-equilibrium
thermodynamics.

Before exploring the uses of these concepts in molecular biology,
let me reiterate the most important points which tend to be
obscured when discussing information. Information is defined as
the amount of correlation between two systems. It measures the
amount of entropy {\em shared} between two systems, and this
shared entropy is the information that one system has {\em about
the other}. Perhaps this is the key insight that I would like to
convey: Information is always {\em about something}. If it cannot
be specified what the information is about, then we are dealing
with entropy, not information. Indeed, entropy is sometimes
called, in what borders on an abuse of language, ``useless
information". The previous discussion also implies that
information is only defined {\it relative} to the system it is
information about, and is therefore {\em never} absolute. This
will be particularly clear in the discussion of the information
content of genomes, which we now enter.

\section{Information in Genomes}
There is a long history of applying information theory to symbolic
sequences. Most of this work is concerned with the randomness, or,
conversely, regularity, of the sequence. Ascertaining the
probabilities with which symbols are found on a sequence or
message will allow us to estimate the entropy of the {\em source
of symbols}, but not what they stand for. In other words,
information cannot be accessed in this manner.  It should be
noted, however, that studying {\em horizontal} correlations, i.e.,
correlations between symbols along a sequence rather than across
sequences, can be useful for distinguishing coding from non-coding
regions in DNA (Grosse et al., 2000), and can serve as a distance
measure between DNA sequences that can be used to assemble
fragments obtained from shotgun-sequencing (Otu and Sayood, 2003).

In terms of the jargon introduced above, measuring the
probabilities with which symbols (or groups of symbols) appear
{\em anywhere} in a sequence will reveal the {\em marginal}
entropy of the sequence, i.e., the entropy {\em without} regard to
the environment or context. The entropy {\em with} regard to the
environment is the entropy {\em given} the environment, a
conditional entropy, which we shall calculate below. This will
involve obtaining the probability to find a symbol at a {\em
specific} point in the sequence, as opposed to anywhere on it. We
sometimes refer to this as obtaining the {\em vertical}
correlations between symbols.

\subsection{Meaning from Correlations}
Obtaining the marginal entropy of a genetic sequence can be quite
involved (in particular if multi-symbol probabilities are
required), but a very good approximative answer can be given
without any work at all: This entropy (for DNA sequences) is about
two bits per base. There are deviations of interest (for example
in GC-rich genes, etc.) but overall this is what the
(non-conditional) entropy of most of DNA is (see, e.g., Schmitt
and Herzel 1997)\nocite{SchmittHerzel1997}. The reason for this is
immediately clear: DNA is a {\em code}, and codes do not reveal
information from sequence alone. Optimal codes, e.g., are such
that the encoded sequences cannot be compressed any further (Cover
and Thomas, 1991)\nocite{CoverThomas1991}. While DNA is not
optimal (there are some correlations between symbols along the
sequence), it is nearly so. The same seems to hold true for
proteins: a random protein would have $\log_2(20)=4.32$ bits of
entropy per site (or 1 mer, the entropy of a random monomer
introduced above), while the actual entropy is somewhat lower due
to biases in the overall abundance (leucine is over three times as
abundant as tyrosine, for example), and due to pair and triplet
correlations. Depending on the data set used, the protein entropy
per site is between 2.5 (Strait and Dewey, 1996) and 4.17 bits
(Weiss et al., 2000)\nocite{StraitDewey1996,Weissetal2000}, or
between 0.6 and 0.97 mers. Indeed, it seems that protein sequences
can only be compressed by about 1\% (Weiss et al. 2000). This is a
pretty good code! But this entropy per symbol only allows us to
quantify our uncertainty about the sequence identity, but it will
not reveal to us the {\em function} of the genes. If this is all
that information theory could do, we would have to agree with the
critics that information theory is nearly useless in molecular
biology. Yet, I have promised that information theory {\em is}
relevant, and I shall presently point out how. First of all, let
us return to the concept of information. How should we decide
whether or not {\em potential information} (a.k.a entropy) is in
{\em actuality} information, i.e., whether it is shared with
another variable?

The key to information lies in its use to make predictions {\em
about} other systems. Only in {\em reference} to another ensemble
can entropy become information, i.e., be promoted from useless to
useful, from potential to actual. Information therefore is clearly
not stored {\em within} a sequence, but rather in the {\em
correlations} between the sequence and what it describes, or what
it {\em corresponds to}. What do biomolecular sequences correspond
to? What is the {\em meaning} of a genomic sequence, what
information does it represent?  This depends, quite naturally, on
what environment the sequence is to be interpreted within.
According to the arguments advanced here, no sequence has an
intrinsic meaning, but only a relative (or conditional) one with
respect to an environment. So, for example, the genome of {\it
Mycoplasma pneumoniae} (a bacterium that causes pneumonia-like
respiratory illnesses) has an entropy of almost a million base
pairs, which is its genome length. Within the soft tissues that it
relies on for survival, most of these base pairs (about 89\%) are
information (Dandekar et al., 2000). Indeed, Mycoplasmas are
obligate parasites in these soft tissues, having shed from 50\% to
three quarters of the genome of their bacterial ancestors (the
{\em Bacillae}). Within these soft tissues that make many
metabolites readily available, what was information for a Bacillus
had become entropy for the Mycoplasma. With respect to {\em other}
environments, the Mycoplasma information might mean very little,
i.e., it might not {\em correspond} to anything there. Whether or
not a sequence means something in its environment determines
whether or not the organism hosting it lives or dies there. This
will allow us to find a way to distinguish entropy from
information in genomes.

\subsection{Physical Complexity}
In practice, how can we determine whether a particular base's
entropy is shared, i.e., whether a nucleotide carries entropy or
information? At first glance one might fear that we would have to
know a gene's function (i.e., know what it corresponds to within
its surrounding) before we can determine the information content;
that, for example, we might need to know that a gene codes for an
alcoholdehydrogenase before we can ascertain which base pairs code
for it. Fortunately, this is not true. What is clear, however, is
that we may never distinguish entropy from information if we are
only given a {\em single} sequence to make this determination,
because, in a single sequence, symbols that carry information are
indistinguishable from those that do not. The trick lies in
studying {\em functionally equivalent sets} of sequences, and the
substitution patterns at each aligned position. In an equilibrated
population, i.e, one where sufficient time has passed since the
last evolutionary innovation or bottleneck, we expect a position
that codes for information to be nearly uniform {\em across} the
population (meaning that the same base pair will be found at that
position in all sequences of that population), because a mutation
at that position would detrimentally affect the fitness of the
bearer, and, over time, be purged from the ensemble (this holds in
its precise form only for asexual populations). Positions that do
not code for information, on the other hand, are selectively
neutral, and, with time, will take on all possible symbols at that
position. Thus, we may think of each position on the genome as a
four-sided die. A priori, the uncertainty (entropy) at each
position is two bits, the maximal entropy:
\begin{equation}
H = -\sum_{i= {\rm G,C,A,T}}p(i)\log_2 p(i)=\log_2 4 = 2 \ \ {\rm bits}
\end{equation}
because, a priori, $p(i)=1/4$. For the {\it actual} entropy, we need the actual
probabilities $p_j(i)$, for each position $j$ on the sequence. In a pool of $N$
sequences, $p_j(i)$ is estimated by counting the number $n_j(i)$ of occurrences of
nucleotide $i$ at position $j$, so that $p_j(i)=n_j(i)/N$. This should be done for
all positions $j=1,...,L$ of the sequence, where $L$ is the sequence length.
Ignoring correlations {\em between} positions $j$ on a sequence (so-called
``epistatic'' correlations, to which we shall return below), the information stored
in the sequence is then (with logs to base 2)

\begin{equation}
I=H_{\rm max}- H = 2L -H\;\; {\rm bits}\;, \label{infomeasure}
\end{equation}
where
\begin{equation}
H = -\sum_{j=1}^L\,\sum_{i={\rm G,C,A,T}}p_j(i)\log_2 p_j(i)\;. \label{cond}
\end{equation}
Note that this estimate, because it relies on the difference of
maximal and actual entropy, does not require us to know which
variables in the environment cause some nucleotides to be uniform,
or ``fixed''. These probabilities are set by mutation-selection
balance in the environment. I have argued earlier (Adami and Cerf
2000, Adami et al. 2000)\nocite{Adamietal2000} that the
information stored in a sequence is a good proxy for the
sequences's complexity (called ``physical complexity"), which
itself might be a good predictor of functional complexity. And
indeed, it seems to correspond to the quantity that increases
during Darwinian evolution (Adami 2002a)\nocite{Adami2002a}. We
will encounter below an evolutionary experiment that seems to
corroborate these notions.

In general (for sequences taken from any monomer alphabet of size $D$), the
information stored in the sequence is
\begin{eqnarray}
I=H_{\rm max}-H &=& L-\left(-\sum_{j=1}^L\,\sum_{i=1}^{D}p_j(i)\log_D p_j(i)\right)\\
&=& L-J\;\; {\rm mers}\;, \label{infomer}
\end{eqnarray}
where $J$ can be thought of as the number of {\it non-functional}
(i.e., ``junk") instructions, and I
remind the reader that we defined the ``mer" as the entropy of a
random monomer, normalized to lie between zero and one.

\subsection{Application to DNA and RNA}
In the simplest case, the environment is essentially given by the
intra-cellular binding proteins, and the measure
(\ref{infomeasure}) can be used to investigate the information
content of DNA binding sites (this use of information theory was
pioneered by Schneider et al., 1986). Here, the sample of
sequences can be provided by a sample of equivalent binding sites
within a single genome.  For example, the latter authors aligned
the sequences of 149 {\it E. coli} and coliphage ribosome binding
sites in order to calculate the substitution probabilities at each
position of a 44 base pair region (which encompasses the 34
positions that can be said to constitute the binding site).
\begin{figure}[h]
\centerline{\psfig{figure=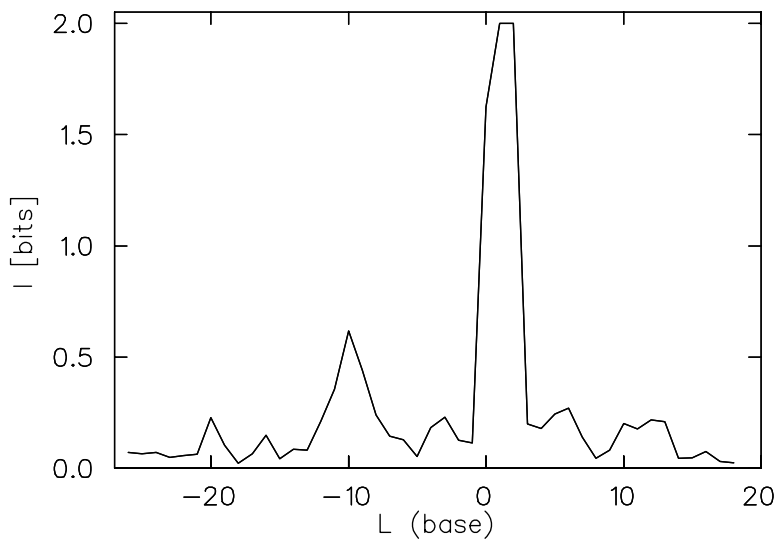,width=4in,angle=90}}
 \caption{Information content (in bits) of an {\it E. coli} ribosome
 binding site, aligned at the $f$Met-tRNA$_f$ initiation site (L=0),
 from Schneider et al.\ (1986).
\label{schneider}}
\end{figure}
Fig. 1 shows the information content as a function of position
(Schneider et al.\ 1986), where position $L=0$ is the first base
of the initiation codon. The information content is highest near
the initiation codon, and shows several distinct peaks. The peak
at $L=-10$ corresponds to the Shine-Dalgarno sequence (Shine and
Dalgarno, 1974).

When the information content of a base is zero we must assume that it
has no function, i.e., it is neither expressed nor does anything
bind to it. Regions with positive information
content\footnote{Finite sampling of the substitution probabilities
introduces a systematic error in the information content, which
can be corrected (Miller 1954, Basharin 1959, Schneider et al.\
1986). In the present case, the correction ensures that the
information content is approximately zero at the left and right
edge of the binding site.} carry information about the binding
protein, just as the binding protein carries information about the
binding site.

It is important to emphasize that the reason that sites $L=1$ and $L=2$, for
example, have maximal information content is a consequence of the fact that their
{\em conditional} entropy Eq.~(\ref{cond}) vanishes.  The entropy is conditional
because only {\em given} the environment of binding proteins in which it functions
in {\em E. coli} or a coliphage, is the entropy zero. If there were, say, two
different proteins which could initiate translation at the same site (two different
environments), the conditional entropy of these sites could be higher. Intermediate
information content (between zero and 2 bits) signals the presence of {\em
polymorphisms} implying either non-specific binding to one protein or competition
between more than one protein for that site.

A polymorphism is a deviation from the consensus sequence that is
not, as a rule, detrimental to the organism carrying it. If it
was, we would call it a ``deleterious mutation'' (or
just``mutation"). The latter should be very infrequent as it
implies disease or death for the carrier. On the contrary,
polymorphisms can establish themselves in the population, leading
either to no change in the phenotype whatsoever, in which case we
may term them ``strictly neutral'', or they may be deleterious by
themselves but neutral if associated with a commensurate
(compensatory) mutation either on the same sequence or somewhere
else.

Polymorphisms are easily detected if we plot the per-site
entropies of a sequence vs.\ residue or nucleotide number in an
{\em entropy map} of a gene. Polymorphisms carry per-site
entropies intermediate between zero (perfectly conserved locus)
and unity (strictly neutral locus). Mutations, on the other hand,
(because they are deleterious) are associated with very low
entropy (Rogan and Schneider 1995), so polymorphisms stand out
among conserved regions and even mutations. In principle,
mutations can occur on sites which are themselves polymorphic;
those can only be detected by a more in-depth analysis of
substitution patterns such as suggested in Schneider (1997).
Because polymorphic sites in proteins are a clue to which sites
can easily be mutated, per-site entropies have also been
calculated for the directed evolution of proteins and enzymes
(Saven and Wolynes 1997, Voigt et al. 2001).

\begin{figure}[!h]
\centerline{\psfig{figure=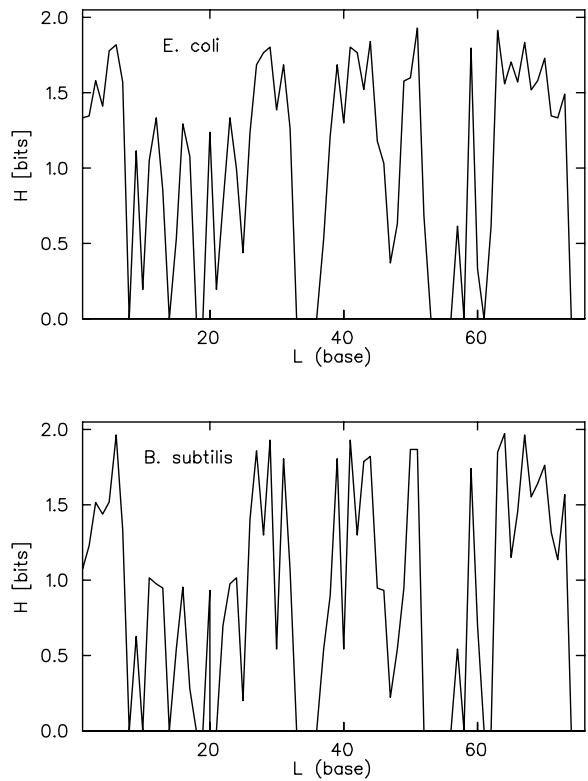,width=3in,angle=0}}
 \caption{Entropy (in bits) of {\it E. coli} tRNA (upper panel)
  from 5' (L=0) to 3' (L=76), from
 33 structurally similar sequences obtained from Sprinzl et al. (1996), where we
 arbitrarily set the entropy of the anti-codon to zero.
Lower panel: Same for 32 sequences of {\it B. subtilis} tRNA. \label{ecolisubt}}
\end{figure}

As mentioned earlier, the actual function of a sequence is irrelevant for
determining its information content. In the previous example, the region
investigated was a binding site. However, any gene's information content can be
measured in such a manner. In Adami and Cerf (2000), the information content of the
76 base pair nucleic acid sequence that codes for bacterial tRNA was investigated.
In this case the analysis is complicated by the fact that the tRNA sequence
displays secondary and tertiary structure, so that the entropy of those sites that
bind in Watson-Crick pairs, for example, are shared, reducing the information
content estimate based on Eq.~(\ref{info}) significantly. In Fig.~\ref{ecolisubt},
I show the entropy (in bits)
\begin{figure}[h]
\centerline{\psfig{figure=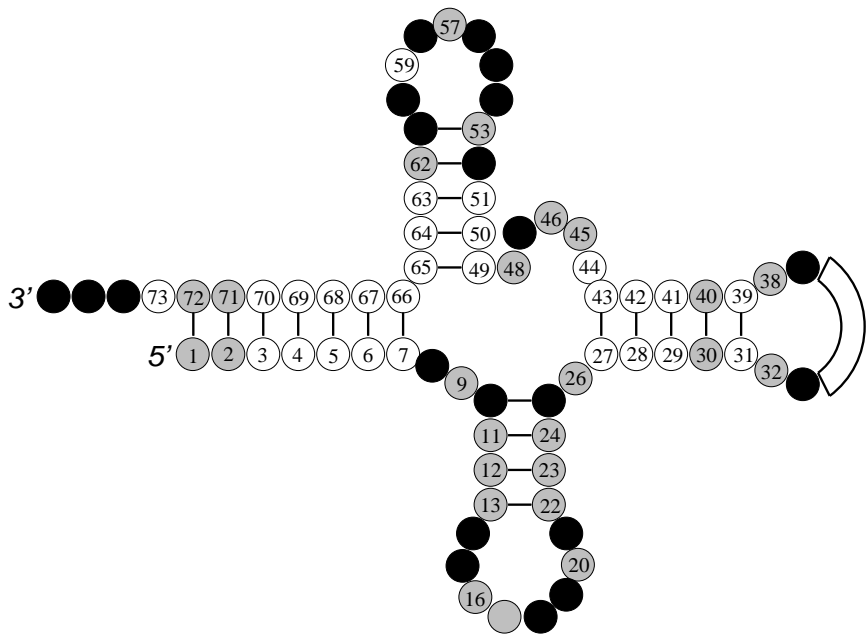,width=4.5in,angle=0}}
\caption{Secondary structure of tRNA molecule, with bases colored
black for low entropy ($0\leq H\leq0.3$ mers), grey for
intermediate ($0.3<H\leq 0.7$ mers), and white for maximal entropy
($0.7<H\leq 1.0$ mers), numbered 1-76 (entropies from {\it E.
coli} sequences). \label{tRNA}}
\end{figure}
derived from 33 structurally similar sequences of {\it E. coli}
tRNA (upper panel) and 32 sequences of {\it B. subtilis} tRNA,
respectively, obtained from the EMBL nucleotide sequence library
(Sprinzl et al. 1996). Note how similar these entropy maps are
across species (even though they last shared an ancestor over 1.6
billion years ago), indicating that the profiles are
characteristic of the {\em function} of the molecule, and thus
statistically stable.

Because of base-pairing, we should not expect to be able to simply
sum up the per-site entropies of the sequence to obtain the
(conditional) sequence entropy. The pairing in the stacks (the
ladder-like arrangement of bases that bind in pairs) of the
secondary structure (see Fig.~\ref{tRNA}) reduces the actual
entropy, because two nucleotides that are bound together {\em
share} their entropy. This is an example where {\em epistatic
correlations} are important. Two sites (loci) are called epistatic
if their contributions to the sequence's fitness are not
independent, in other words, if the probability to find a
particular base at one position depends on the identity of a base
at another position. Watson-Crick-binding in stacks is the
simplest such example; it is also a typical example of the
maintenance of polymorphisms in a population because of functional
association. Indeed, the fact that polymorphisms are correlated in
stacks makes it possible to deduce the secondary structure of an
RNA molecule from sequence information alone.
\begin{figure}[h]
\centerline{\psfig{figure=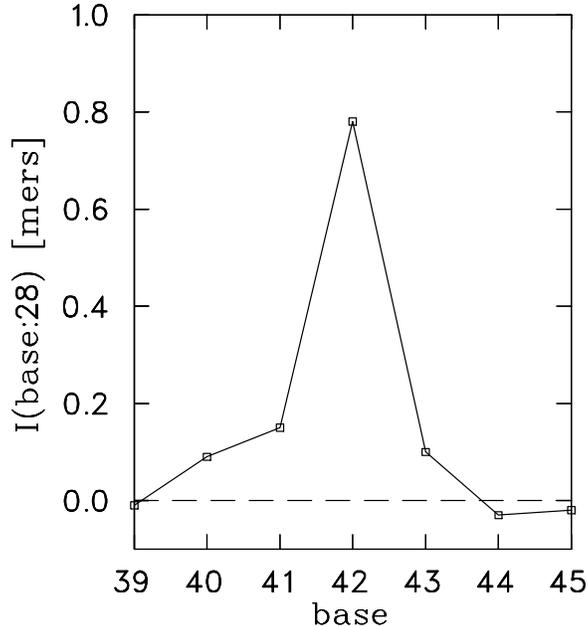,width=3in,angle=0}}
\caption{Mutual
 entropy (information) between base 28 and bases 39 to 45 (information
 is normalized to $I_{\rm max}=1$ by taking logarithms to base 4). Because
 finite sample size corrections of higher order have been neglected, the
 information estimate can appear to be negative by an amount of the order of this
 error.  \label{spike}}
\end{figure}
Take, for example, nucleotide $L=28$ (in the anti-codon stack) which is bound to
nucleotide $L=42$, and let us measure entropies in mers (by taking logarithms to
the base 4). The mutual entropy between $L=28$ and $L=42$ (in {\em E. coli}) can be
calculated using Eq. (4):
\begin{equation}
I(28:42)=H(28)+H(42)-H(28,42)=0.78\;. \label{info28}
\end{equation}
Thus indeed, these two bases share almost all of their entropy. We
can see furthermore that they share very little entropy with any
other base. Note that, in order to estimate the entropies in
Eq.~(\ref{info28}), we applied a first-order correction that takes
into account a bias due to the finite size of the sample, as
described in Miller (1954). This correction amounts to $\Delta
H_1=3/(132 \ln2)$ for single nucleotide entropies, and $\Delta
H_2=15/(132 \ln2)$ for the joint entropy. In Fig.~\ref{spike}, I
plot the mutual entropy of base 28 with bases 39 to 45
respectively, showing that base 42 is picked out unambiguously.
Such an analysis can be carried out for all pairs of nucleotides,
so that the secondary structure of the molecule is revealed
unambiguously (see, e.g., Durbin et al. 1998). In
Fig.~\ref{wagenaar}, I show the entropy (in bits) for all pairs of
bases of the set of {\it E. coli} sequences used to produce the
entropy map in Fig.~\ref{ecolisubt}, which demonstrates how the
paired bases in the four stems stand out.
\begin{figure}[h]
\centerline{\psfig{figure=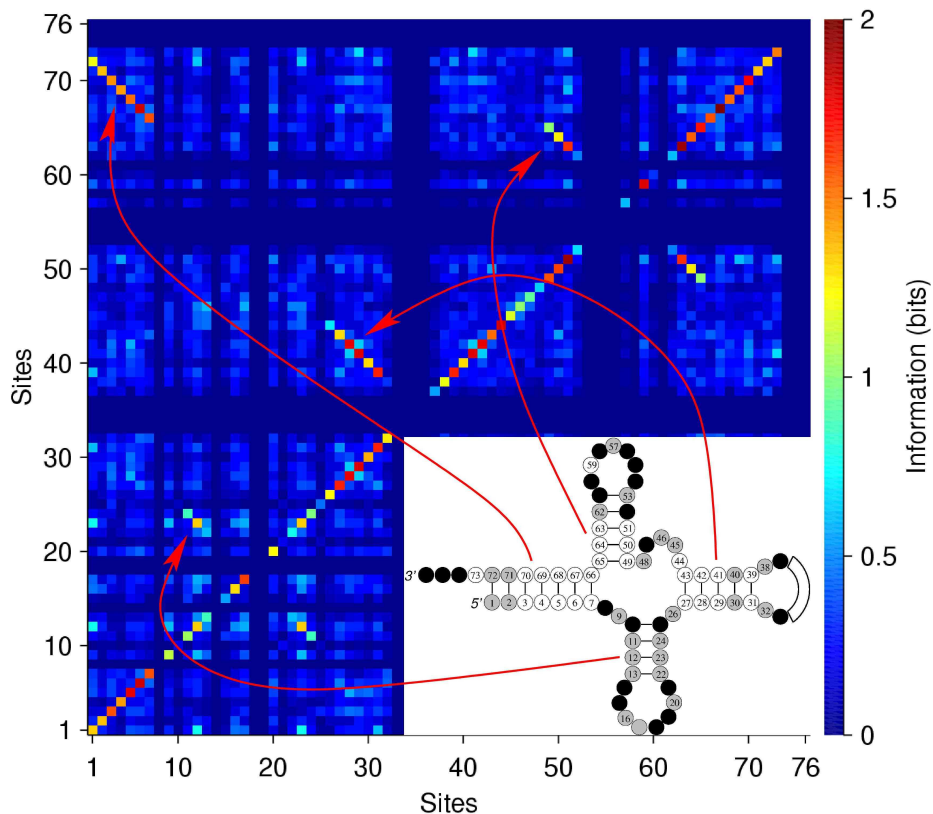,width=5in,angle=0}}
\caption{Mutual
 entropy (information) between all bases (in bits), colored according
 to the color bar on the right, from 33 sequences of {\it E. coli} tRNA.
 The four stems are readily identified by their correlations as indicated.
  \label{wagenaar}}
\end{figure}

Since we found that most bases in stacks share almost all of their
entropy with their binding partner, it is easy to correct formula
(\ref{infomer}) to account for the epistatic effects of
stack-binding: We only need to subtract from the total length of
the molecule (in mers) the number of bases involved in stack
binding. In a tRNA molecule (with a secondary structure as in
Fig.~\ref{tRNA}) there are 21 such bases, so the sum in
Eq.~(\ref{cond}) should only go over the 52 ``reference
positions"\footnote{We exclude the three anticodon-specifying
bases from the entropy calculation because they have zero
conditional entropy by {\em definition} (they cannot vary among a
tRNA-type because it would change the type). However, the
substitution probabilities are obtained from mixtures of {\em
different} tRNA-types, and therefore appear to deviate from zero or one.}.
For {\it E. coli}, the entropy summed over the reference positions
gives $H\approx 24$ mers, while the {\it B. subtilis} set gives
$H\approx 21$ mers. We thus conclude that bacterial tRNA stores
between 52 and 55 mers of information about its environment (104-110
bits).

This type of sequence analysis combining structural and complexity
information has recently been used to quantify the information
gain during in-vitro evolution of catalytic RNA molecules
(ribozyme ligases) (Carothers et al.\
2004)\nocite{Carothersetal2004}. The authors evolved RNA aptamers
that bind GTP (guanine triphosphate) with different catalytic
effectiveness (different functional capacity) from a mutagenized
sequence library. They found 11 different classes of ribozymes,
whose structure they determined using the correlation analysis
outlined above. They subsequently measured the amount of
information in each structure [using Eq.~(\ref{infomeasure}) and
correcting for stack binding as described above] and showed that
ligases with higher affinity for the substrate had more complex
secondary structure {\em and} stored more information.
Furthermore, they found that the information estimate based on
Eq.~(\ref{infomeasure}) was consistent with an interpretation in
terms of the amount of information necessary to specify the
particular structure in the given environment. Thus, at least in
this restricted biochemical example, structural, functional, and
informational complexity seem to go hand in hand.

\subsection{Application to Proteins}
If the secondary structure of RNA and DNA enzymes can be predicted
based on correlations alone, what about protein secondary
structure? Because proteins fold and function via the interactions
among the amino acids they are made of, these interactions should,
in evolutionary time, lead to correlations between residues so
that the fitness effect of an amino acid substitution at one
position will depend on the residue at another position. (Care
must be taken to avoid contamination from correlations that are
due entirely to a common evolutionary path, see Wollenberg and
Atchley 2000; Govindarajan et al.\
2003.)\nocite{WollenbergAtchley2000,Govindarajanetal2003} Such an
analysis has been carried out on a number of different molecule
families, such as the V3 loop region of HIV-1 (Korber et al.
1993)\nocite{Korberetal1993}, which shows high variability (high
entropy) and strong correlations between residues (leading to
shared entropy) that are due to functional constraints. These
correlations have also been modelled (Giraud et al.\
1998)\nocite{Giraudetal1998}.

A similar analysis for the homeodomain sequence family was
performed by Clarke (1995)\nocite{Clarke1995}, who was able to
detect 16 strongly co-varying pairs in this 60 amino acid binding
motif. However, determining secondary structure based on these
correlations alone is much more difficult, because proteins do not
fold neatly into stacks and loops as does RNA. Also, residue
covariation does not necessarily indicate physical proximity
(Clarke 1995)\nocite{Clarke1995}, even though the strongest
correlations are often due to salt-bridges. But the correlations
can at least help in eliminating some models of protein structure
(Clarke 1995).

Atchley et al.\ (2000)\nocite{Atchleyetal2000} carried out a
detailed analysis of correlated mutations in the bHLH (basic
helix-loop-helix) protein domain of a family of transcription
factors. Their set covered 242 proteins across a large number of
vertebrates that could be aligned to detect covariation. They
found that amino acid sites known to pack against each other
showed low entropy, whereas exposed non-contact sites exhibited
significantly larger entropy. Furthermore, they determined that a
significant amount of the observed correlations between sites was
due to functional or structural constraints that could help in
elucidating the structural, functional, and evolutionary dynamics
of these proteins (Atchley et al. 2000).

Some attempts have been made to study the {\em thermodynamics} of protein
structures and relate it to the sequence entropy (Dewey 1997)\nocite{Dewey1997}, by
studying the mutual entropy between protein sequence and {\em structure}. This line
of thought is inspired by our concept of the genotype-phenotype map, which implies
that sequence should predict structure. If we hypothesize a structural entropy of
proteins $H({\rm str})$, obtained for example as the logarithm of the possible
stable protein structures for a given chain length (and a given environment), then
we can write down the mutual entropy between structure and sequence simply as
\begin{equation}
I({\rm seq}:{\rm str})=H({\rm seq})-H({\rm seq|str})\;, \label{seqstr}
\end{equation}
where $H({\rm seq})$ is the entropy of sequences of length $L$, given by $L$, and
$H({\rm seq|str})$ is the entropy of sequences {\em given} the structure. If we
assume that the environment perfectly dictates structure (i.e., if we assume that
only one particular structure will perform any given function) then
\begin{equation}
H({\rm seq|str})\approx H({\rm seq}|{\rm env})
\end{equation}
and $I({\rm str}:{\rm seq})$ is then roughly equal to the physical
complexity defined earlier. Because $H({\rm str|seq})=0$ (per the
above assumption that any given sequence gives rise to exactly one
structure), we can rewrite (\ref{seqstr}) as
\begin{equation}
I({\rm seq}:{\rm env})\approx I({\rm seq}:{\rm str})=H({\rm
str})-\underbrace{H({\rm str}|{\rm seq})}_{=0} \;,
\end{equation}
i.e., the mutual entropy between sequence and structure only tells
us that the thermodynamical entropy of possible protein structures
is limited by the amount of information about the environment
coded for by the sequence. This is interesting because it implies
that sequences that encode more information about the environment
are also potentially more complex, a relationship we discussed
earlier in connection with ribozymes (Carothers et al.\ 2004).
Note, however, that the assumption that only one particular
structure will perform any given function need not hold. Szostak
(2003), for example, advocates a definition of {\em functional
information} that allows for different structures carrying out an
equivalent biochemical function.

\section{Molecular Interactions and Resistance}

One of the more pressing concerns in bioinformatics is the identification of DNA
protein-binding regions, such as promoters, regulatory regions, and splice
junctions. The common method to find such regions is through {\em sequence
identity}, i.e., known promoter or binding sites are compared to the region being
scanned (e.g., via freely available bioinformatics software such as BLAST), and a
``hit'' results if the scanned region is sufficiently identical according to a
user-specified threshold. Such a method cannot, of course, find {\em unknown}
binding sites, nor can it detect interactions between proteins, which is another
one of bioinformatics' holy grails (see, e.g., Tucker et al.
2001)\nocite{TuckerGeraUetz2001}. Information theory can in principle detect
interactions between different molecules (such as DNA-protein or protein-protein
interactions) from {\it sequence heterogeneity}, because interacting pairs share
{\em correlated mutations}, that arise as follows.

\subsection{Detecting Protein-Protein and DNA-Protein Interactions}
Imagine two proteins bound to each other, while each protein has
some entropy in its binding motif (substitutions that do not
affect structure). If a mutation in one of the proteins leads to
severely reduced interaction specificity, the substitution is
strongly selected against. It is possible, however, that a {\em
compensatory} mutation in the binding partner restores
specificity, such that the {\em pair} of mutations together is
neutral (and will persist in the population), while each mutation
by itself is deleterious. Over evolutionary time, such pairs of
correlated mutations will establish themselves in populations and
in homologous genes across species, and could be used to identify
interacting pairs. This effect has been seen previously in the
Cytochrome c/Cytochrome oxidase (CYC/COX) heterodimer (Rawson and
Burton 2002)\nocite{RawsonBurton2002} of the marine copepod {\it
Tigriopus californicus}. In Rawson and Burton (2002), the authors performed crosses between
the San Diego (SD) and Santa Cruz (SC) variants from two natural
allopatric populations that have long, independent evolutionary
histories. Inter-population crosses produced strongly reduced
activity of the cytochrome complex, while intra-population crosses
were vigorous. Indeed, the SD and SC variants of COX differ by at
least 30 amino acid substitutions, while the smaller CYC has up to
5 substitutions. But can these correlated mutations be found from
sequence data alone? This turns out to be a difficult
computational problem unless it is known precisely which member of
a set of $N$ sequences of one binding partner binds to which
member of a set of $N$ of the other. Unless we are in possession
of this $N$ to $N$ assignment, we cannot calculate the joint
probabilities $p_{ij}$ that go into the calculation of the mutual
entropies such as Eq.~(\ref{info28}) that reveal correlated
mutations.

Of course, if we have one pair of sequences from $N$ species of organisms with the
same homologous gene, the assignment is automatically implied. In the absence of
such an assignment, it may be possible to recover the correct matches from two sets
of $N$ sequences by searching for the assignment with the highest mutual entropy,
because we can safely assume that the correct assignment maximizes the correlations
(Adami and Thomsen 2004)\nocite{AdamiThomsen2004}. However, this is a difficult
search problem because the number of possible assignments scales like $N!$. Still,
because correlated mutations due to coevolution seem to be relatively common
(Bonneton et al. 2003)\nocite{Bonnetonetal2003}, this would be a useful tool for
revealing those residues involved in binding, or even in protein-protein
interaction prediction.

In principle, the information-theoretical method described above
can potentially identify {\em unknown} binding sites by
identifying complementary patterns (between binding sites and
protein coding regions), if the binding regions are not
well-conserved, i.e., when the binding site and the corresponding
transcription factor carry a reasonable amount of polymorphisms,
and if enough annotation exists to identify the genomic partners
that correspond to each other in a set. If sufficient pairs of
transcription-factor/binding-domain pairs can be sequenced, an
information-theoretic analysis could conceivably reveal genomic
regulatory regions that standard sequence analysis methods miss.
For example, it was suggested recently (Brown and Callan,
2004)\nocite{BrownCallan2004} that the cAMP response protein (CRP,
a transcription factor that regulates many {\it E. coli} genes)
binds to a number of entropic sites in {\it E.coli}, i.e., sites
that are not strictly conserved, but that still retain
functionality (see also Berg and von Hippel, 1987).

\begin{figure}[h]
\centerline{\psfig{figure=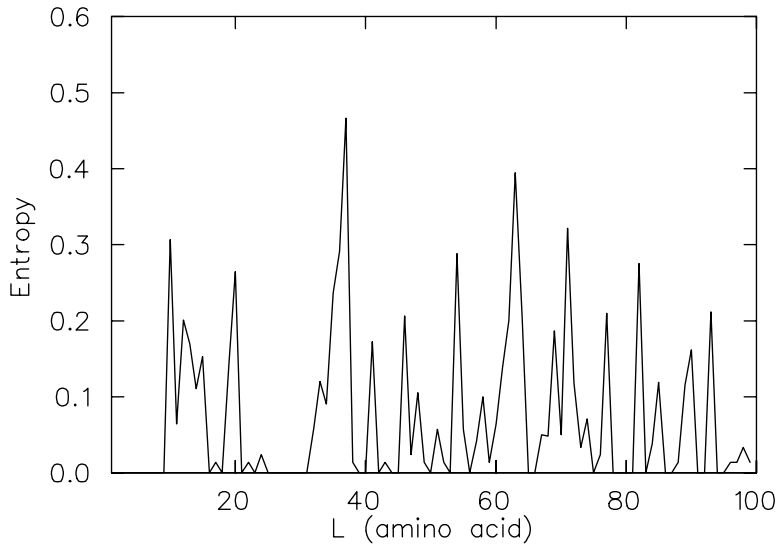,width=4in,angle=90}}
\caption{Normalized ($0\le H\le1$) entropy of HIV-1 protease in
mers, as a function of residue number, using 146 sequences from
patients exposed to a protease inhibitor drug (entropy is
normalized to $H_{\rm max}=1$ per amino acid by taking logarithms
to base 20).\label{hivent}}
\end{figure}

\subsection{Tracking Drug Resistance}

An interesting pattern of mutations can be observed in the
protease of HIV-1, a protein that binds to particular motifs on a
virus polyprotein, and then cuts it into functional pieces.
Resistance to protease inhibitors (small molecules designed to
bind to the ``business end" of the protease, thereby preventing
its function) occurs via mutations in the protease that do not
change the protease's cutting function (proteolysis), while
preventing the inhibitor to bind to it. Information theory can be
used to study whether mutations are involved in drug resistance or
whether they are purely neutral, and to discover correlated
resistance mutations.

The emergence of resistance mutations in the protease after
exposure to antiviral drugs has been well studied (Molla et al.
1996, Schinazi, Larder, and Mellors 1999). The entropy map of HIV
protease in Fig.~\ref{hivent}\footnote{The map was created using
146 sequences obtained from a cohort in Luxembourg, and deposited
in GenBank (Servais et al.\ 1999 and 2001a).}  (on the level of
amino acids) reveals a distinctive pattern of polymorphisms and
only two strictly conserved regions. HIV protease {\em not}
exposed to inhibitory drugs, on the other hand, shows three such
conserved regions (Loeb et al. 1989)\nocite{Loebetal1989}. It is
believed that the polymorphisms contribute to resistance mutations
involved in HAART (Highly Active Antiretroviral Therapy) failure
patients (Servais et al.\ 2001b). But, as a matter of fact, many
of the observed polymorphisms can be observed in treatment-naive
patients (Kozal et al. 1996, and Lech et al. 1996) so it is not
immediately clear which of the polymorphic sites are involved in
drug resistance.

\begin{figure}[h] \centerline{\psfig{figure=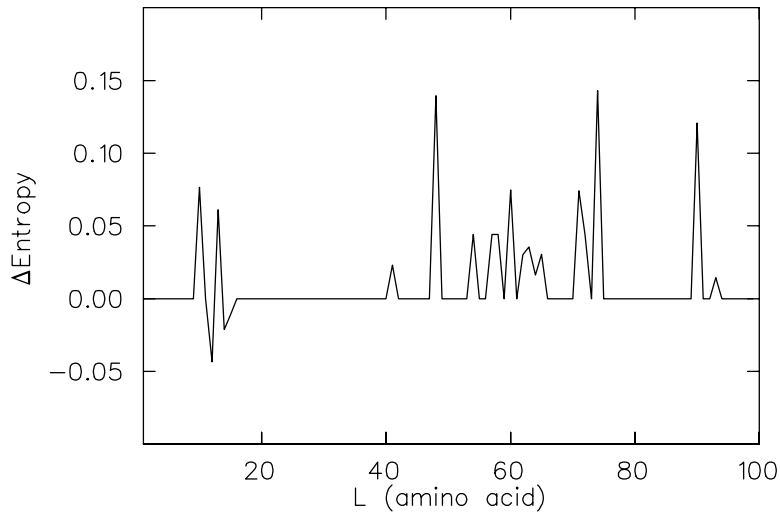,width=4in,angle=90}}
\caption{Change in per-site entropy of HIV-1 protease after six
months of exposure to saquinavir, $\Delta$Entropy=$H_{26}-H_0$,
where $H_{26}$ is the entropy after 26 weeks of exposure. The
entropies were obtained from 34 sequences before and after
exposure, available through GenBank (Schapiro et al.\ 1997). The
three highest (positive) spikes are associated to the well-known
resistance mutations G48V, T74(A,S), and L90M, respectively.
}\label{figdiff}
\end{figure}

In principle, exposure of a population to a new environment can
lead to fast adaptation if the mutation rate is high enough. This
is certainly the case with HIV. The adaptive changes generally
fall into two classes: mutations in regions that were previously
conserved (true resistance mutations), and changes in the
substitution pattern on sites that were previously polymorphic. In
the case of HIV-1 protease, both patterns seem to contribute. In
Fig.~\ref{figdiff}, I show the {\it changes} in the entropic
profile of HIV-1 protease obtained from a group of patients before
and six months after treatment with high doses of saquinavir (a
protease inhibitor). Most spikes are positive, in particular the
changes around residues 46-56, a region that is well-conserved in
treatment-naive proteases, and that is associated with a {\em
flap} in the molecule that must be flexible and that extends over
the substrate binding cleft (Shao et al. 1997). Mutations in that
region indeed appeared on sites that were previously uniform,
while some changes occurred on polymorphic sites (negative
spikes). For those, exposure to the new environment usually
reduced the entropy at that site.

Some of the resistance mutations actually appear in pairs,
indicating that they may be compensatory in nature (Leigh Brown et
al. 1999, Hoffman et al. 2003, Wu et al. 2003)\nocite{Wuetal2003}.
The strongest association occurs between residues 54 and 82, the
former associated with the flap, and the latter right within the
binding cleft. This association does not occur in treatment-naive
patients, but stands out strongly after therapy (such correlations
are easily detected by creating mutual entropy graphs such as
Fig.~\ref{wagenaar}, data not shown). The common explanation for
this covariation is again compensation: while a mutation in the
flap or in the cleft leads to reduced functionality of the
protease, both together restore function while evading the
inhibitor.

\subsection{Information-theoretic Drug Design}
Because many of the protease polymorphisms are prevalent in
treatment-naive patients, we must assume that they are either
neutral, or that the steric changes they entail do not impede the
protease's proteolytic activity while failing to bind the protease
inhibitor. Thus, a typical protease population is a mixture of
polymorphic molecules (polymorphic both in genotype and in
structure, see Maggio et al. 2002)\nocite{Maggioetal2002} that can
outsmart a drug designed for a single protease type relatively
easily. An interesting alternative in drug design would therefore
use an entropic mixture of polymorphisms, or ``quasispecies''
(Eigen 1971) as the drug target. Such a drug would {\em itself}
form a quasispecies rather than a pure drug. Indeed, an analysis
of the information content of realistic ensembles shows that
consensus sequences are exceedingly rare in real populations
(Schneider 1997), and certainly absent in highly variable ones
such as HIV proteases. The absence of a consensus sequence is also
predicted for molecules evolving at the {\em error threshold}
(Eigen 1971), which is very likely in these viruses.

The ideal {\em superdrug} should represent a mixture of inhibitors
that is perfectly tuned to the mixture of proteases. What this
mixture is can be determined with information theory, by ensuring
that the ensemble of inhibitors {\em co-varies} with the protease,
such as to produce tight binding even in the presence of mutations
(or more precisely {\em because} of the presence of mutations).
The substitution probabilities of the inhibitor ensemble would be
obtained by maximizing the mutual entropy (information) between
the protease and an inhibitor library obtained by combinatorial
methods, either on a nucleotide or on the amino acid level (Adami
2002b)\nocite{Adami2002b}. If such a procedure could create a drug
that successfully inhibits resistance mutations, we could no
longer doubt the utility of information theory for molecular
biology.

\section{Conclusions}
Information theory is not widely used in bioinformatics today even
though, as the name suggests, it should be {\em the} relevant
theory for investigating the information content of sequences. The
reason for the neglect appears to be a misunderstanding of the
concepts of entropy versus information throughout most of the
literature, which has led to the widespread perception of its
incompetence. Instead, I point out that Shannon's theory precisely
defines both entropy and information, and that our intuitive
concept of information coincides with the mathematical notion.
Using these concepts, it is possible in principle to distinguish
information-coding regions from random ones in ensembles of
genomes, and thus quantify the information content. A thorough
application of this program should resolve the C-paradox, that is,
the absence of a correlation between the size of the genome and
the apparent complexity of an organism (Cavalier-Smith 1985), by
distinguishing information that contributes to complexity from
non-functional stretches that do not. However, this is a challenge
for the future because of the dearth of multiply sequenced
genomes.

Another possible application of information theory in molecular
biology is the association of regulatory molecules with their
binding sites or even protein-protein interactions, in the case
where transcription factors and their corresponding binding site
show a good amount of polymorphism (methods based on correlated
heterogeneity), and the binding association between pairs can be
established. This approach is complementary to sequence comparison
of conserved regions (methods based on sequence identity), in
which information theory methods cannot be used because zero
(conditional) entropy regions cannot share entropy. Conversely,
sequence comparison methods must fail if polymorphisms are too
pronounced. Finally, the recognition of the polymorphic (or
quasispecies) nature of many viral proteins suggests an
information theory based approach to drug design in which the
quasispecies of proteins---rather than the consensus sequence---is
the drug target, by maximizing the information shared between the
target and drug ensembles.

\vskip 0.25cm \noindent{\bf Acknowledgements} \vskip 0.25cm I
thank David Baltimore and Alan Perelson for discussions, and Titus
Brown for comments on the manuscript. Thanks are also due to
Daniel Wagenaar for producing Fig.~5. This work was supported by the
National Science Foundation under grant DEB-9981397.

\end{document}